\documentclass[12pt]{article}

\usepackage{scicite}

\usepackage{times}

\usepackage{graphicx}
\usepackage{hyperref}
\usepackage{color}
\newcommand{\wzy}[1]{\textcolor{black}{#1}}

\newenvironment{sciabstract}{%
\begin{quote} \bf}
{\end{quote}}

\title{Breaking the quantum adiabatic speed limit by jumping along geodesics} 

\author
{Kebiao Xu,$^{1,2,4\ast}$ Tianyu Xie,$^{1,2,4\ast}$ Fazhan Shi,$^{1,2,4}$ Zhen-Yu Wang,$^{3\dag}$ 
\\
Xiangkun Xu,$^{1,2,4\ddag}$ Pengfei Wang,$^{1,4}$ Ya Wang,$^{1,4}$ Martin B. Plenio,$^{3}$ Jiangfeng Du,$^{1,2,4\dag}$\\
\\
\normalsize{$^{1}$CAS Key Laboratory of Microscale Magnetic Resonance and Department of Modern Physics,}\\
	\normalsize{University of Science and Technology of China (USTC), Hefei 230026, China}\\
\normalsize{$^{2}$Hefei National Laboratory for Physical Sciences at the Microscale, USTC, Hefei 230026, China}\\
\normalsize{$^{3}$Institut f\"ur Theoretische Physik und IQST, Albert-Einstein-Allee 11,}\\
	\normalsize{Universit\"at Ulm, D-89081 Ulm, Germany}\\
\normalsize{$^{4}$Synergetic Innovation Center of Quantum Information and Quantum Physics,}\\
	\normalsize{USTC, Hefei 230026, China}\\
\\
\normalsize{$^\ast$These authors contributed equally to this work.}\\
\normalsize{$^\dag$Corresponding authors. E-mail:  zhenyu.wang@uni-ulm.de (Z.Y.W); djf@ustc.edu.cn (J.D.).}\\
\normalsize{$^\ddag$Current address: Department of Radiation Oncology,}\\
	\normalsize{Johns Hopkins School of Medicine, Baltimore, MD, 21287, USA}
}

\date{}

\begin{document} 


\baselineskip24pt


\maketitle 

\textbf{One Sentence Summary: } Conventional restriction on adiabatic techniques can be removed to develop new strategies to control quantum systems.

\begin{sciabstract}
Quantum adiabatic evolutions find a broad range of applications in quantum physics and quantum technologies. The traditional form of the quantum adiabatic theorem limits the speed of adiabatic evolution by the minimal energy gaps of the system Hamiltonian. Here we show experimentally using an nitrogen-vacancy center in diamond that even in the presence of vanishing energy gaps quantum adiabatic evolution is possible. This verifies a recently derived necessary and sufficient quantum adiabatic theorem and offers paths to overcome the conventionally assumed constraints on adiabatic methods. By fast modulation of dynamic phases, we demonstrate near unit-fidelity quantum adiabatic processes in finite times. These results challenge traditional views and provide deeper understanding on quantum adiabatic processes as well as promising strategies for the control of quantum systems.
\end{sciabstract}

\section*{Main Text}
\subsection*{Introduction}
Coherent control on quantum systems is a fundamental element of quantum
technologies which could revolutionize the fields of information processing, simulation, and sensing.  A powerful and universal method
to achieve this control is the quantum adiabatic technique, which exhibits intrinsic robustness against control
errors ensured by the quantum adiabatic evolution~\cite{Childs2001}.
Besides important applications in quantum state engineering~\cite{RMP2017, Vitanov2001}, quantum simulation~\cite{Aspuru2005, kim2010, biamonte2011adiabatic}, and quantum computation~\cite{farhi2000quantum, jones2000geometric, Farhi2001, Barends2016,Xu2017}, the quantum adiabatic evolution itself also provides interesting
properties such as Abelian~\cite{berry1984quantal} or non-Abelian geometric phases~\cite{wilczek1984appearance},
which can be used for the realization of quantum gates.
However, the conventional quantum adiabatic theorem~\cite{Born1928, MessiahBook}, 
which dates back to the idea of extremely slow and reversible change in classical mechanics~\cite{Laidler1994, Born1928}, 
imposes a speed limit on the quantum adiabatic methods, that is, for a quantum process to remain adiabatic the changes to the system Hamiltonian at all times must be much smaller than the energy gap of the Hamiltonian.
On the other hand, in order to avoid perturbations from the environment high rates of change are desirable. This tension can impose severe limitations on the practical use of adiabatic methods.
Despite the long history and broad applicability, it was
discovered recently that key aspects of quantum adiabatic evolution remain not fully understood~\cite{marzlin2004inconsistency, tong2005quantitative}
and the condition in the conventional adiabatic theorem is not necessary for quantum adiabatic evolution~\cite{Du2008, wang2016necessary}.

In this work, we experimentally demonstrate adiabatic evolutions with vanishing energy gaps and energy level crossings, which however are allowed in a recently proven quantum adiabatic condition~\cite{wang2016necessary} that is based on dynamical phases instead of energy gaps, by using an NV center~\cite{doherty2013nitrogen} in diamond. 
In addition, we reveal that employing discrete jumps along the evolution path allows quantum adiabatic processes at unlimited rates which challenges the view that adiabatic processes must be slow. By jumping along the path one can even avoid path points where the eigenstates of the Hamiltonian are not feasible in experiments. Furthermore, we demonstrate theoretically and experimentally the elimination of all the non-adiabatic effects on system evolution of a finite evolution time by driving the system along the geodesic that connect initial and final states, as well as combating system decoherence by incorporating pulse sequences into adiabatic driving.

\subsection*{Experimental study of the necessary  and sufficient quantum adiabatic condition}
To describe the theory for experiments, consider a quantum system driven
by a Hamiltonian $H(\lambda)$ for adiabatic evolution. In terms of its instantaneous orthonormal eigenstates $|\psi_{n}(\lambda)\rangle$
($n=1,2,\ldots$) and eigenenergies $E_{n}(\lambda)$, the Hamiltonian is written as $H(\lambda)=\sum_{n}E_{n}(\lambda)|\psi_{n}(\lambda)\rangle\langle\psi_{n}(\lambda)|$. For a given continuous finite evolution path, $|\psi_{n}(\lambda)\rangle$ changes gradually with the configuration parameter $\lambda$. In our experiments, $\lambda$ corresponds to an angle in some unit and is tuned in time  such that $\lambda=\lambda(t)\in[0,1]$. The system dynamics driven by the Hamiltonian
is fully determined by the corresponding evolution propagator $U(\lambda)$.
It is shown that one can decompose the propagator $U(\lambda)=U_{\rm{adia}}(\lambda) U_{\rm{dia}}(\lambda)$ as the product of a quantum adiabatic evolution propagator $U_{\rm{adia}}(\lambda)$ that describes the ideal quantum evolution in the adiabatic limit and a diabatic propagator $U_{\rm{dia}}(\lambda)$ that includes \emph{all} the diabatic errors~\cite{wang2016necessary}. In the adiabatic limit, $U_{\rm{dia}}(\lambda)=I$ becomes an identity matrix and the adiabatic evolution
$U=U_{\rm{adia}}(\lambda)$ fully describes the geometric phases~\cite{berry1984quantal,wilczek1984appearance} and dynamic phases accompanying the adiabatic evolution (that is, the deviation from adiabaticity  $U-U_{\rm{adia}}$ vanishes). This decomposition guarantees that both $U_{\rm{adia}}(\lambda)$ and $U_{\rm{dia}}(\lambda)$ are gauge invariant, i.e., invariant with respect to any chosen state basis.

According to the result of~\cite{wang2016necessary}, the error part
satisfies the first-order differential equation ($\hbar=1$),
\begin{equation}
\frac{d}{d\lambda}U_{\rm{dia}}(\lambda)=iW(\lambda)U_{\rm{dia}}(\lambda),\label{eq:Udia}
\end{equation}
with the boundary condition $U_{\rm{dia}}(0)=I$. The generator $W(\lambda)$ describes \emph{all} the non-adiabatic transitions.
In the basis of $|\psi_{n}(0)\rangle$, the diagonal matrix elements of $W(\lambda)$
vanishes, i.e., $\langle\psi_{n}(0)|W(\lambda)|\psi_{n}(0)\rangle=0$.
The off-diagonal matrix elements
\begin{equation}
\langle\psi_{n}(0)|W(\lambda)|\psi_{m}(0)\rangle=e^{i\phi_{n,m}(\lambda)}G_{n,m}(\lambda)\label{eq:W}
\end{equation}
are responsible for non-adiabaticity. Here $\phi_{n,m}(\lambda)\equiv\phi_{n}(\lambda)-\phi_{m}(\lambda)$ is the
difference of the accumulated dynamic phases $\phi_{n}(\lambda)$ on
$|\psi_{n}(\lambda)\rangle$, and
the geometric part $G_{n,m}(\lambda)=e^{i\left[\gamma_{m}(\lambda)-\gamma_{n}(\lambda)\right]}g_{n,m}(\lambda)$ consists of the geometric functions $g_{n,m}(\lambda)=i\langle\psi_{n}(\lambda)|\frac{d}{d\lambda}|\psi_{m}(\lambda)\rangle$ and the geometric phases $\gamma_{n}(\lambda)=\int_{0}^{\lambda}g_{n,n}(\lambda^{\prime})d\lambda^{\prime}$.

Equation~\ref{eq:W} show that the differences
of dynamic phases $\phi_{n,m}$ are more fundamental than the energy gaps in suppressing
the non-adiabatic effects, because the energies $E_{n}$ do not explicitly appear in these equations.
Indeed, according to~\cite{wang2016necessary}, when the  dynamic phase factors at different path points add destructively
\begin{equation}
\epsilon_{n,m}(\lambda)=\left|\int_{0}^{\lambda}e^{i\phi_{n,m}(\lambda^{\prime})}d\lambda^{\prime}\right|<\epsilon, \label{eq:avg}
\end{equation}
for \wzy{$n\neq m$ and} any $\lambda\in[0,1]$ of a finite path with bounded $G_{n,m}(\lambda)$ and $\frac{d}{d\lambda}G_{n,m}(\lambda)$, the deviation from adiabaticity
can be made arbitrarily small by reducing $\epsilon$ with a scaling factor determined by
the magnitudes of $G_{n,m}(\lambda)$ and $\frac{d}{d\lambda}G_{n,m}(\lambda)$. 
That is, the operator norm $||U_{\rm{dia}}(\lambda)-I||<\sqrt{\epsilon}(G_{\rm{tot}}^2+G_{\rm{tot}}^\prime)\lambda^2+(\sqrt{\epsilon}+\epsilon) G_{\rm{tot}}$, where
$G_{\rm{tot}}=\sum_{n\neq m}{\rm{max}}|G_{n,m}(\lambda^\prime)|$ and $G_{\rm{tot}}^\prime=\sum_{n\neq m}{\rm{max}}|\frac{d}{d\lambda^\prime}G_{n,m}(\lambda^\prime)|$ for $0<\lambda^\prime\leq\lambda$~\cite{wang2016necessary}.
In the limit $\epsilon\rightarrow 0$
the system evolution is adiabatic along the entire finite path with $U_{\rm{dia}}(\lambda)\rightarrow I$.
For a zero gap throughout the evolution path, the evolution is not adiabatic because $\epsilon_{n,m}(\lambda)=\lambda$ is not negligible due to the constructive interference of the dynamic phase factors at different path points. For a large constant gap, the destructive interference gives a negligible $\epsilon_{n,m}$ and hence an adiabatic evolution.

\begin{figure*}
\center
\includegraphics[width=1.0\textwidth]{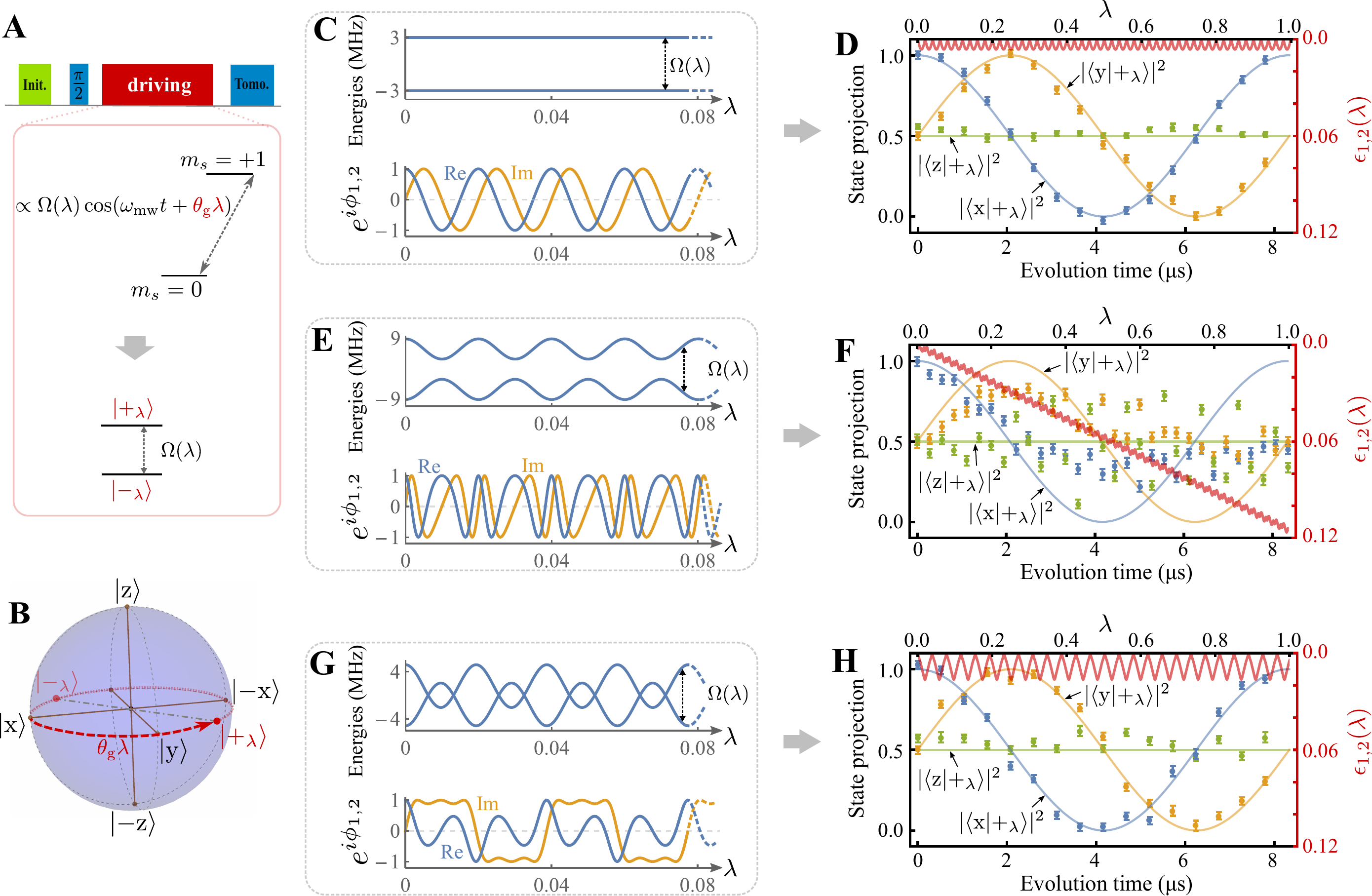}
\caption{\label{fig1:FigConstPi}\textbf{Quantum adiabaticity of continuous driving}. (\textbf{A}) Illustration of experimental control. A microwave resonant with an NV transition forms in the rotating frame a Hamiltonian with the instantaneous eigenstates $|\psi_{1}(\lambda)\rangle=|+_{\lambda}\rangle$ and $|\psi_{2}(\lambda)\rangle=|-_{\lambda}\rangle$ that are separated by an energy gap $\Omega(\lambda)$ proportional to the amplitude of the microwave field. 
(\textbf{B}) Evolving path (red curve with an arrow head) on the Bloch sphere when increasing the parameter $\lambda=\lambda(t)$ in the microwave phase with the evolution time $t$. 
(\textbf{C}) The energies of the eigenstates for a constant gap $\Omega(\lambda)=\Omega_{0}=2\pi\times6$ MHz and the corresponding real and imaginary parts of $e^{i\phi_{1,2}}$ as a function of $\lambda$. 
(\textbf{D}) The measured projections (dots) of the system state \wzy{on the $|\rm{x}\rangle$, $|\rm{y}\rangle$, and $|\rm{z}\rangle$ states. The system state} was initialized in $|+_{\lambda=0}\rangle=|\rm{x}\rangle$, see (B), and subsequently driven by the Hamiltonian with the eigenenergies shown in (C) and with \wzy{a} changing rate $d\lambda/dt=0.12$~MHz \wzy{and} a path length $\theta_{\rm{g}}=2\pi$\wzy{. T}he lines show the ideal state projections of the instantaneous eigenstate $|+_{\lambda}\rangle$. The red line is a plot of $\epsilon_{1,2}(\lambda)$, i.e., the interference of $e^{i\phi_{1,2}}$ at different path points.
(\textbf{E} and \textbf{F}) Same as (C) and (D), respectively, but for a gap $\Omega(\lambda)=\Omega_{0}[2+\cos(\Omega_{0} \lambda T)]$ larger than the gap in (C). Because $\epsilon_{1,2}(\lambda)$ is not negligible, the corresponding evolution in (F) is not adiabatic.
(\textbf{G} and \textbf{H}) Same as (C) and (D), respectively, but for the gap $\Omega(\lambda)=\Omega_{\pi}(\lambda)$ that has energy level crossings. $\epsilon_{1,2}(\lambda)$ is negligible and induces the quantum adiabatic evolution shown in (H). The error bars in all the figures represent two standard errors of the mean.
}
\end{figure*}

To experimentally verify the adiabatic condition \ref{eq:avg} by an NV center,
we construct the Hamiltonian for adiabatic evolution in the standard way~\cite{RMP2017,Vitanov2001}.
That is, we apply a microwave field to drive the NV electron spin states
$|m_{\rm{s}}=0\rangle\equiv|-{\rm{z}}\rangle$ and $|m_{\rm{s}}=+1\rangle\equiv |{\rm{z}}\rangle$ (see Fig.~\ref{fig1:FigConstPi}A, \wzy{Materials and Methods} for experimental details). 
The Hamiltonian $H(\lambda)$ under an on-resonant microwave field reads
\begin{equation}
H_{\rm{XY}}(\lambda) = \frac{\Omega(\lambda)}{2}\Big [ |\psi_{1}(\lambda)\rangle\langle\psi_{1}(\lambda)|-|\psi_{2}(\lambda)\rangle\langle\psi_{2}(\lambda)| \Big ], \label{eq:Hxy}
\end{equation}
where the energy gap $\Omega(\lambda)$ is tunable, 
and the instantaneous eigenstates of the system Hamiltonian $|\psi_{1}(\lambda)\rangle=|+_{\lambda}\rangle$ and $|\psi_{2}(\lambda)\rangle=|-_{\lambda}\rangle$. Here
\begin{equation}
|\pm_{\lambda}\rangle\equiv\frac{1}{\sqrt{2}}(|{\rm{z}}\rangle\pm e^{i\theta_{\rm{g}}\lambda}|-{\rm{z}}\rangle) \label{eq:xyPath}
\end{equation}
are tunable by varying the microwave phases $\theta_{\rm g}\lambda$. We define the initial eigenstates $|\pm\rm{x}\rangle\equiv|\pm_{0}\rangle$ and the superposition states $|\pm{\rm{y}}\rangle\equiv\frac{1}{\sqrt{2}}({|{\rm{z}}\rangle} \pm i {|-{\rm{z}}\rangle})$ for convenience.

In the traditional approach that the Hamiltonian varies slowly with a non-vanishing
gap, the strength of relative dynamic phase \wzy{$\phi_{1,2}=\phi_{1}(\lambda)-\phi_{2}(\lambda)$}
rapidly increases with the change of the path parameter $\lambda$, giving
the fast oscillating factor $e^{i\phi_{1,2}}$ 
with a zero mean (see Fig.~\ref{fig1:FigConstPi}C for the case of a constant gap $\Omega(\lambda)=\Omega_{0}$). 
Therefore the right-hand side of 
Eq.~\ref{eq:Udia} is negligible in solving the differential equation, leading to
the solution $U_{\rm{dia}}(\lambda)\approx I$. As a consequence of the adiabatic evolution $U\approx U_{\rm{adia}}(\lambda)$,
the state initialized in an initial eigenstate of the Hamiltonian follows the evolution of the instantaneous eigenstate (see Fig.~\ref{fig1:FigConstPi}D).

\begin{figure*}\center
\includegraphics[width=1.0\textwidth]{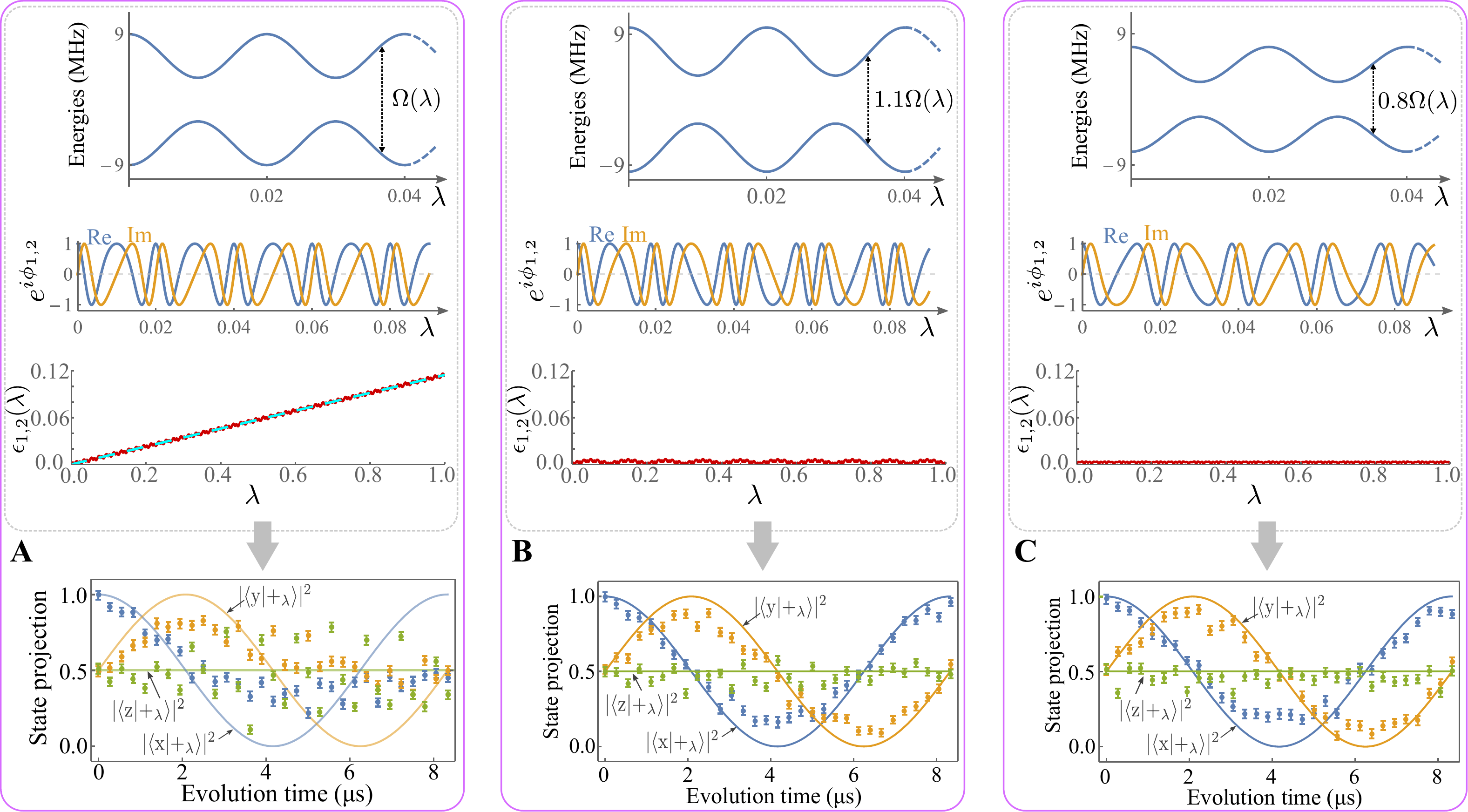}
\caption{\label{fig2:FigSDia} \textbf{Recovery of quantum adiabaticity by adding energy gap fluctuations.} 
(\textbf{A}) The energies of instantaneous eigenstates, the real and imaginary parts of $e^{i\phi_{1,2}}$, $\epsilon_{1,2}(\lambda)$, and the measured projections (dots) of the system state on the $|\rm{x}\rangle$, $|\rm{y}\rangle$, or $|\rm{z}\rangle$ states. The results are the same as those in Fig.~\ref{fig1:FigConstPi} (E and F) where the energy gap $\Omega(\lambda)=\Omega_{0}[2+\cos(\Omega_{0} \lambda T)]$. 
(\textbf{B}) \wzy{[(\textbf{C})]} Same as (A) but changing the energy gap as $\Omega(\lambda)\rightarrow 1.1\Omega(\lambda)$ \wzy{[}$\Omega(\lambda)\rightarrow 0.8\Omega(\lambda)$\wzy{],} by adding an amplitude bias in the control field of the experiments. The fluctuation in the energy gap induces random modulation on the function $e^{i\phi_{1,2}}$. The destructive interference on $e^{i\phi_{1,2}}$ leads to a smaller average $\epsilon_{1,2}(\lambda)$ and hence improved quantum adiabatic evolution. \wzy{In (A) the} cyan dashed line in the plot \wzy{of} $\epsilon_{n,m}(\lambda)$ shows the line $J_{2}(1)\lambda\approx 0.115 \lambda$.
}
\end{figure*}

However, a quantum evolution with a non-vanishing gap and a long evolution time is not necessary adiabatic. In Fig.~\ref{fig1:FigConstPi} (E and F)
we show a counterexample that increasing the energy gap in Fig.~\ref{fig1:FigConstPi}C to $\Omega(\lambda)=\Omega_{0}[2+\cos(\Omega_{0} \lambda T)]\geq \Omega_{0}$ 
will not realize adiabatic evolution because in this case the $\epsilon_{1,2}(\lambda)$ in Eq.~\ref{eq:avg} and the $G_{1,2}(\lambda)$ are not negligible.
For example, $\epsilon_{1,2}(\lambda)=J_{2}(1)\lambda \approx 0.115\lambda$ ($J_{n}$ being the Bessel function of the first kind) 
whenever the difference of dynamic phases is a multiple of $2\pi$. This counterexample is different from the previously proposed counterexamples~\cite{marzlin2004inconsistency,tong2005quantitative,
wang2016necessary,Ortigoso2012} where the Hamiltonian contains resonant terms which increase $|\frac{d}{d\lambda}G_{n,m}(\lambda)|$ and hence modify the evolution path when increasing the total time. Our
counterexample also demonstrates that the widely used adiabatic condition~\cite{MessiahBook} $|\langle \psi_{n}| \frac{d}{dt}|\psi_{m}\rangle|/|E_n-E_m|\ll 1$, 
which is based on the energy gap and diverges at $E_n-E_m=0$, does not guarantee quantum adiabatic evolution. On the contrary, the condition Eq.~\ref{eq:avg} based on dynamic phases, i.e., integrated energy differences, does not diverge for any energy gaps.
We note that fast amplitude fluctuations on the control fields (hence energy gaps) can exist in adiabatic methods [e.g., see \cite{Jing2014}] because of their strong robustness against control errors. Indeed, by adding errors in the energy gap shown in Fig.~\ref{fig1:FigConstPi}E,
the adiabaticity of the evolution is significantly enhanced (see Fig.~\ref{fig2:FigSDia}), showing that the situations to have 
non-adiabatic evolution with a fluctuating energy gap are relatively rare.

We demonstrate that adiabatic evolution can be achieved even when the energy spectrum exhibits vanishing gaps and crossings as long as Eq.~\ref{eq:avg} is satisfied for a sufficiently small $\epsilon$. As an example,
we consider the energy gap of the form $\Omega(\lambda)=\Omega_{\pi}(\lambda)\equiv\Omega_{0}^{\prime}\left[1+a\cos(2\Omega_{0}^{\prime} T \lambda)\right]$, which has zeros and crossings for $|a|>1$ (see Fig.~\ref{fig1:FigConstPi}G for the case of $a\approx 2.34$, where $\Omega_{0}^{\prime}=\sqrt{2/(2+a^2)}\Omega_0$
is used to have the same average microwave power in both Fig.~\ref{fig1:FigConstPi}C and Fig.~\ref{fig1:FigConstPi}G). 
Despite the vanishing gaps and crossings, the corresponding factor $e^{i\phi_{1,2}}$ parameterized by the parameter $\lambda=t/T$ is fast oscillating (see Fig.~\ref{fig1:FigConstPi}G) with a zero mean and realizes quantum adiabatic evolution for a sufficiently large total time $T$ (see Fig.~\ref{fig1:FigConstPi}H). In
%fig.~\ref{fig:FigSGapChange}
fig.~S1, we show how the adiabaticity can also be preserved when gradually introducing energy level crossings.

\subsection*{Unit-fidelity quantum adiabatic evolution within a finite time}
\begin{figure*}
\center
\includegraphics[width=0.72\textwidth]{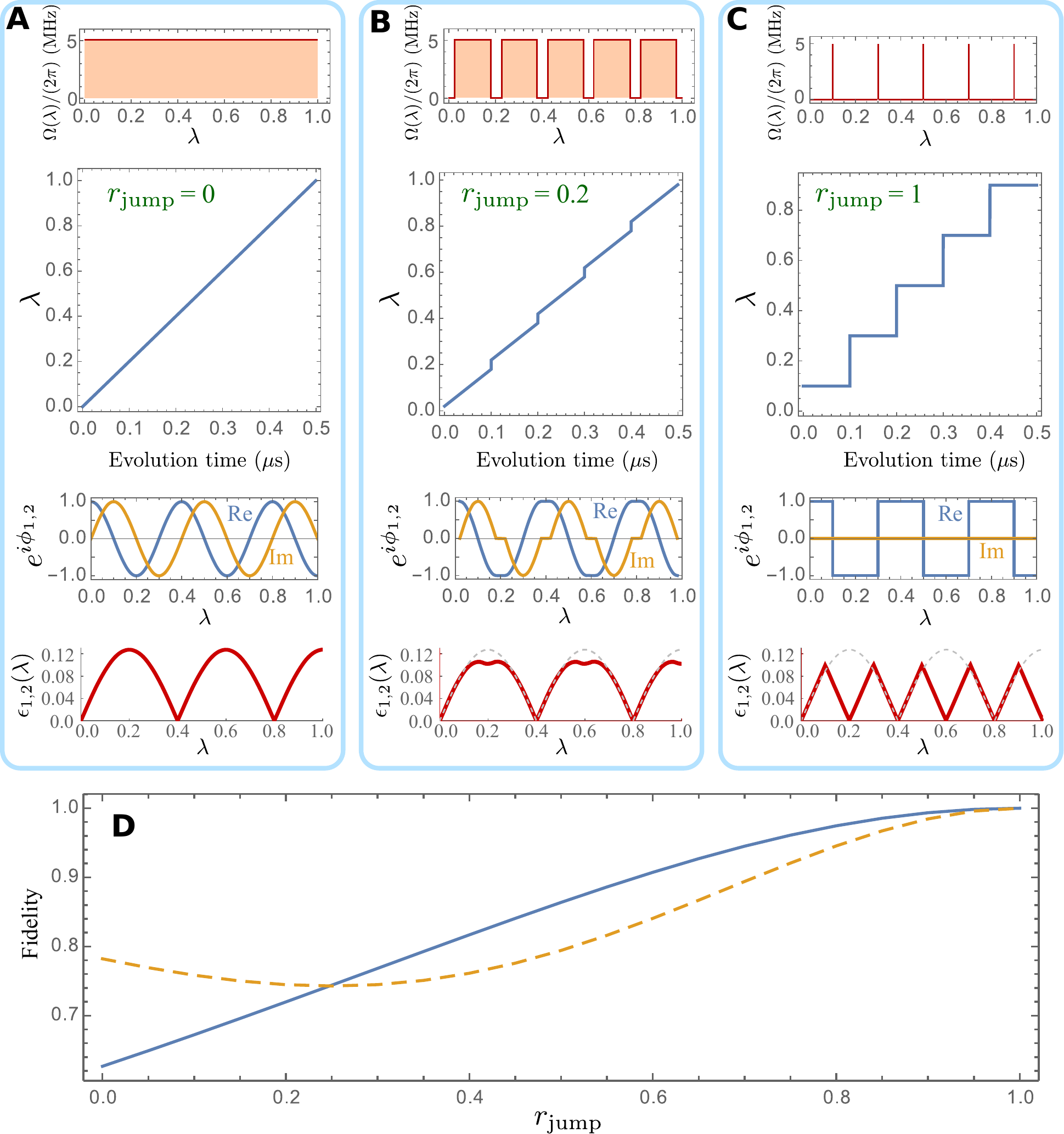}
\caption{\label{fig3:Figc2jump}\textbf{Transition of adiabatic driving from the standard continuous protocol to the jumping protocol}. (\textbf{A}) Relations among the energy gap $\Omega(\lambda)$, path parameter $\lambda$, evolution time $t$, the phase factor $e^{i\phi_{1,2}(\lambda)}$, and $\epsilon_{1,2}(\lambda)$ for the standard adiabatic driving with a constant gap $\Omega_{0}=2\pi\times5~$MHz. (\textbf{B} and \textbf{C}) Same as (A) with the maximum gap $\Omega_{0}=2\pi\times5~$MHz but with a ratio $r_{\rm{jump}}$ of intervals to have $\Omega(\lambda)=0$ along the path parameter $\lambda$. Therefore, the case of $r_{\rm{jump}}=0$ corresponds to the standard adiabatic protocol without a vanishing gap. A ratio $r_{\rm{jump}}>0$ in (B) opens regions that have $\Omega(\lambda)=0$. For the maximum value $r_{\rm{jump}}=1$ we get in (C) the jumping protocol which only drives the system at discrete path points with a Rabi frequency \wzy{equalling $\Omega_{0}$}. For comparison, the plot of $\epsilon_{1,2}(\lambda)$ in (A) is also shown in (B) and (C) by a gray dashed line. (\textbf{D}) Calculated fidelity to the ideal adiabatic state at the final time as a function of $r_{\rm{jump}}$.
The fidelity increases to 100\% when the driving is getting to the jumping protocol. The solid line shows the case that the initial state is prepared in the initial eigenstate $|\rm{x}\rangle$ of the Hamiltonian, while the dashed line is the result for the initial \wzy{state being the} superposition state $|\rm{y}\rangle$. \wzy{The driving is along the adiabatic path given by Eq.~(\ref{eq:xyPath}) with $\theta_{\rm g}=\pi$ and is repeated back and forth three times for a total time $T=3$ $\mu$s.}}
\end{figure*}

Without the restriction to non-zero energy gaps, it is possible to
completely eliminate non-adiabatic effects and to drive an arbitrary initial
state $|\Psi_{\rm{i}}\rangle$ to a target state $|\Psi_{\rm{t}}\rangle$ of a general quantum system
by the quantum adiabatic evolution of a finite time duration. We demonstrate this
by driving the system along the geodesic
for maximal speed [see, e.g., \cite{anandan1990geometry,chruscinski2004geometric} for more discussion on the geodesic in quantum mechanics]. The system eigenstate
$|\psi_{1}(\lambda)\rangle=\cos(\frac{1}{2}\theta_{\rm{g}}\lambda)|\psi_{1}(0)\rangle+\sin(\frac{1}{2}\theta_{\rm{g}}\lambda)|\psi_{2}(0)\rangle$
connects $|\psi_{1}(0)\rangle$ and $|\psi_{1}(1)\rangle$ along the geodesic by varying
$\lambda=0$ to $\lambda=1$, with its orthonormal eigenstate
$|\psi_{2}(\lambda)\rangle=-\sin(\frac{1}{2}\theta_{\rm{g}}\lambda)|\psi_{1}(0)\rangle+\cos(\frac{1}{2}\theta_{\rm{g}}\lambda)|\psi_{2}(0)\rangle$
varied accordingly  \wzy{(see Materials and Methods)}. The method works for any quantum system \wzy{(e.g., a set of interacting qubits)} because a geodesics can always be found~ \cite{anandan1990geometry,chruscinski2004geometric}.  An example of the geodesic path for a single qubit is given by Eq.~\ref{eq:xyPath}, which intuitively can be illustrated by the shortest path on the Bloch sphere (see Fig.~\ref{fig1:FigConstPi}B). We find that along the geodesic the only nonzero
elements $g_{2,1}(\lambda)$ and $g_{1,2}(\lambda)$ are constant.
We adopt the sequence theoretically proposed in \cite{wang2016necessary}
that changes the dynamic phases at $N$ equally spaced path
points $\lambda=\lambda_{j}$ ($j=1,2,\ldots,N$). By staying
at each of the points
\begin{equation}
\lambda_{j}=(2N)^{-1}(2j-1),
\end{equation}
for a time required to implement a $\pi$ phase shift on the dynamic phases,
we have $U_{\rm{dia}}(1)=I$ because $W(\lambda)$ commutes and $$\epsilon_{1,2}(1)=\left|\int_{0}^{1}e^{i\phi_{1,2}(\lambda)}d\lambda\right|=0.$$
That is, by jumping on discrete points $\lambda_{j}$ the system evolution at $\lambda=1$ is exactly the perfect adiabatic evolution $U_{\rm{adia}}$ even through the evolution time is finite,
and an initial state $|\Psi_{\rm{i}}\rangle$ will end up with the adiabatic target state
$|\Psi_{\rm{t}}\rangle=U_{\rm{adia}}|\Psi_{\rm{i}}\rangle$.
To realize the jumping protocol, we apply rectangular $\pi$ pulses at the points $\lambda_{j}$ without time delay between the pulses because between the points $\lambda_{j}$ the Hamiltonian has a zero energy gap and its driving can be neglected (see Fig.~\ref{fig3:Figc2jump}\wzy{C}). The simulation results in Fig.~\ref{fig3:Figc2jump} show how the transition from the standard continuous protocol to the jumping one gradually increases the fidelity of adiabatic evolution.

\begin{figure*}
\center
\includegraphics[width=1.0\textwidth]{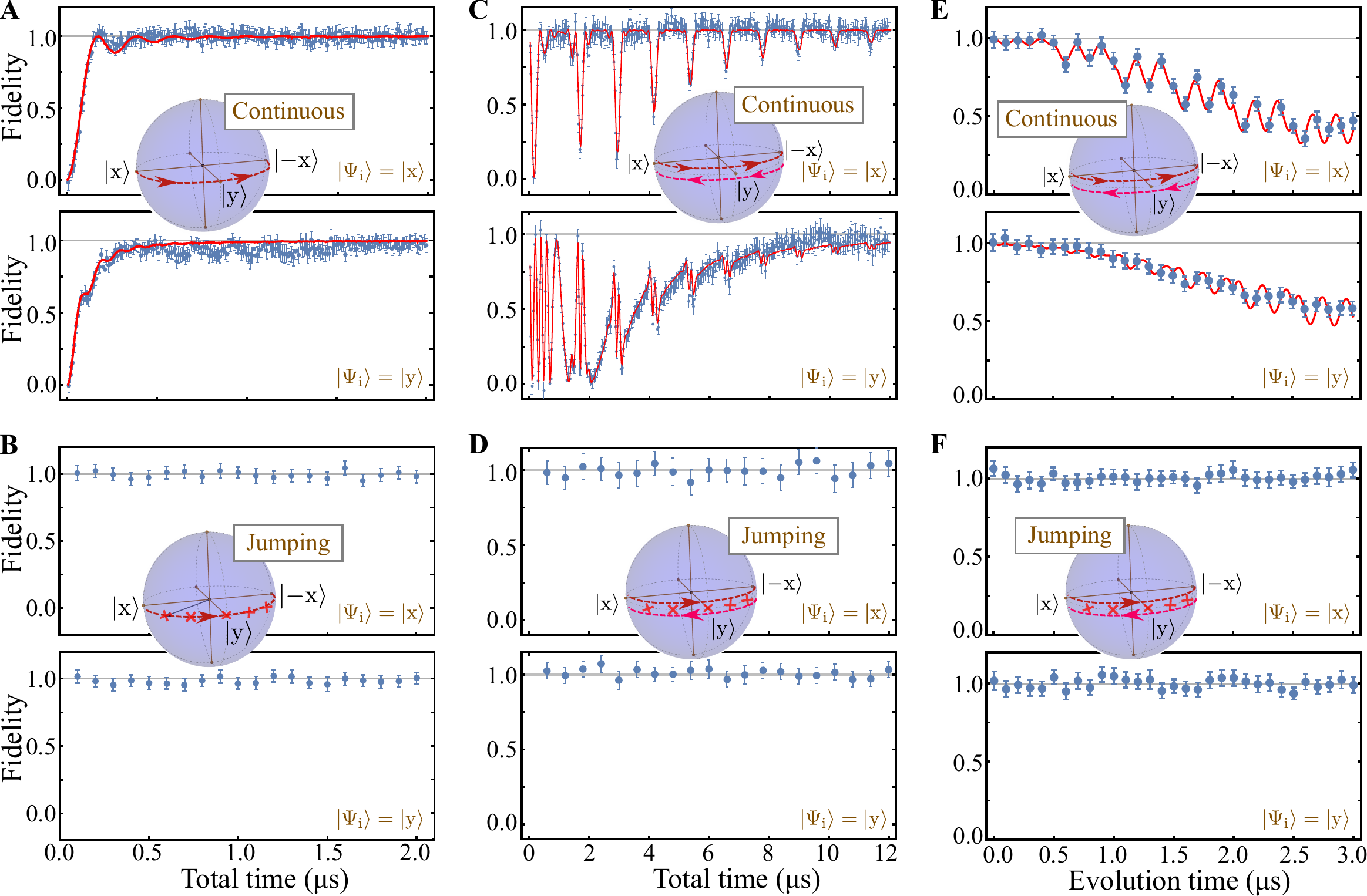}
\caption{\label{fig4:FigPulse}
\textbf{Performance of adiabatic protocols along geodesics.} \wzy{(\textbf{A}) Measured fidelity (blue dots) between the final state and the target adiabatic state as a function of the total time  by using a continuous driving protocol with a constant gap, $\Omega_{0}=2\pi\times 5$ MHz. The inset indicates the control path where the instantaneous eigenstate $|\psi_{1}(\lambda)\rangle$ of the Hamiltonian proceeds from $|\rm{x}\rangle$ to the $|-\rm{x}\rangle$ via a geodesic half circle. The initial state $|\psi_{1}(0)\rangle=|\Psi_{\rm i}\rangle$ is prepared in the initial eigenstate $|\rm{x}\rangle$ (upper panel) or $|\rm{y}\rangle$ (lower panel), the equal superposition of initial eigenstates. The target state is defined as the state driven by ideal, infinitely slow adiabatic evolution. The red lines show the numerical simulation which has taken experimental noise sources into consideration  (see Materials and Methods). The horizontal grey lines indicate the level of unit fidelity. (\textbf{B}) Same as (A) but for a jumping protocol where the $\pi$ pulses have the same amplitude as the continuous protocol but are applied at the path points $\lambda_j$ without time delay. The crosses in the inset illustrate the path points for $N=5$ pulses. (\textbf{C} and \textbf{D}), Same as (A) and (B), respectively, but for a longer path containing six half circles by using three times of back-forward motion. (\textbf{E} and \textbf{F}) Fidelity during the evolution for a total time $T= 3$~$\mu$s using the protocols in (C) and (D).}}
\end{figure*}

\wzy{We experimentally compare} the jumping protocol with the continuous one \wzy{along the geodesic given by Eq.~\ref{eq:xyPath}, by measuring} the fidelity of the evolved state to the target state $|\Psi_{\rm{t}}\rangle$ that follows the ideal adiabatic evolution $U_{\rm{adia}}$. \wzy{The continuous protocol has a constant gap and a constant sweeping rate as in Fig.~\ref{fig1:FigConstPi}C. As shown in Fig.~\ref{fig4:FigPulse} (A and B) for the case of a geodesic half circle ($\theta_{\rm g} =\pi$), the} jumping protocol reaches unit fidelity within the measurement accuracy, while the standard continuous driving has much lower fidelity at short evolution times. \wzy{The advantage of the jumping protocol is more prominent when
we traverse the half-circle path back and forth, see Fig.~\ref{fig4:FigPulse} (C to F) for the results of a total path length of $6\theta_{\rm{g}}$.}
We observe in Fig.~\ref{fig4:FigPulse} that the constant-gap
protocol provides unit state-transfer fidelity only when the initial state is an eigenstate of the initial Hamiltonian $|\Psi_{\rm{i}}\rangle=|\rm{x}\rangle$
and the relative dynamic phase accumulated in a single half circle is $\phi=\sqrt{(2k\pi)^{2}-(\theta_{\rm{g}})^{2}}$ ($k=1,2,\ldots$) \wzy{(see Materials and Methods)}.
However, the phase shifts on the system eigenstates accompanying adiabatic evolution can not be observed
when the initial state is prepared in one of the initial eigenstates. Therefore, in Fig.~\ref{fig4:FigPulse} we
also compare the fidelity \wzy{for the} initial state $|\Psi_{\rm{i}}\rangle=|\rm{y}\rangle$, which is a superposition of the initial eigenstates $|\pm\rm{x}\rangle$. The results confirm that the jumping protocol achieves exactly the adiabatic evolution $U_{\rm{adia}}$ within the experimental uncertainties.

\subsection*{Robustness of quantum adiabatic evolution via jumping}
\begin{figure}\center
\includegraphics[width=0.618\columnwidth]{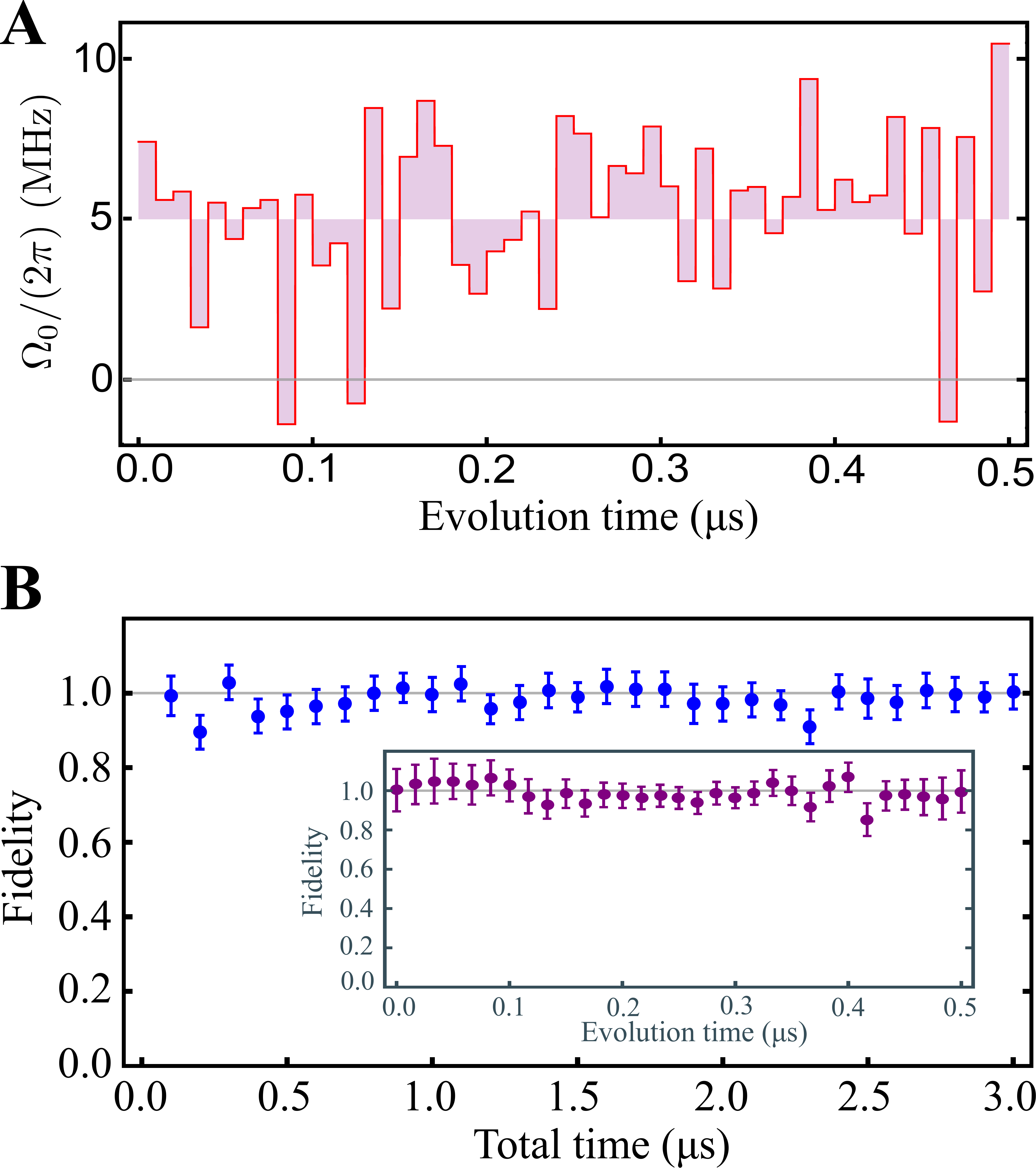}
\caption{\label{fig5:FigRobust}\textbf{Robustness of the jumping protocol.} 
(\textbf{A}) Exemplary time trace of the driving Rabi frequency. The amplitudes of the Rabi frequency are randomly generated by the Gaussian distribution with a mean of $2\pi\times5$ MHz and a standard deviation of $2\pi\times2.5$ MHz. The amplitudes are uncorrelated at every slices of duration of 10~ns.
(\textbf{B}) Fidelity to the final adiabatic state as a function of the total time using the amplitudes as (A) and the initial state prepared
in the eigenstate $|\rm{x}\rangle$. The inset of (B) shows the fidelity during the evolution time $t$ for $N=5$.
The fidelity is measured by comparing the experimental state with the ideal state under infinitely slow adiabatic evolution.}
\end{figure}

To demonstrate the intrinsic robustness guaranteed by adiabatic evolutions,
in Fig.~\ref{fig5:FigRobust}, we consider large random driving amplitude
errors in the jumping protocol. We add random Gaussian distributed errors with a standard
deviation of 50 \% to the control amplitude. To simulate white noise,
we change the amplitude after every 10 ns in an uncorrelated
manner. Despite the large amplitude errors,
which can even cause energy level crossings, during the evolution (see Fig.~\ref{fig5:FigRobust}A for a random time trace), a change of fidelity is hardly observable in Fig.~\ref{fig5:FigRobust}B.
Additional simulations in fig.~S2
%fig.~\ref{fig:Noise}
also demonstrate the robustness to amplitude fluctuations with different kinds of noise correlation, i.e., Gaussian white noise, Ornstein-Uhlenbeck process modeled noise, and static random noise.
The robustness of the jumping protocol can be further enhanced by using a larger number $N$ of points along the path (see fig.~S3).
%fig.~\ref{fig:NoisePN}).

While it is different from dynamical decoupling (DD)~\cite{yang2011preserving,wang2016necessary}, the jumping protocol can suppress the effect of environmental noise through a mechanism similar to DD. Therefore the fidelity is still high even when the evolution time is much longer than the coherence time, $T_{2}^{*}=1.7$~$\mu$s, of the NV electron spin (see fig.~S4).
% fig.~\ref{fig:LongTime}).
This evidence is useful to design adiabatic protocols that provide strong robustness against both control errors and general environmental perturbations.

\subsection*{Avoiding unwanted path points in adiabatic evolution}
\begin{figure}
\center
\includegraphics[width=0.618\columnwidth]{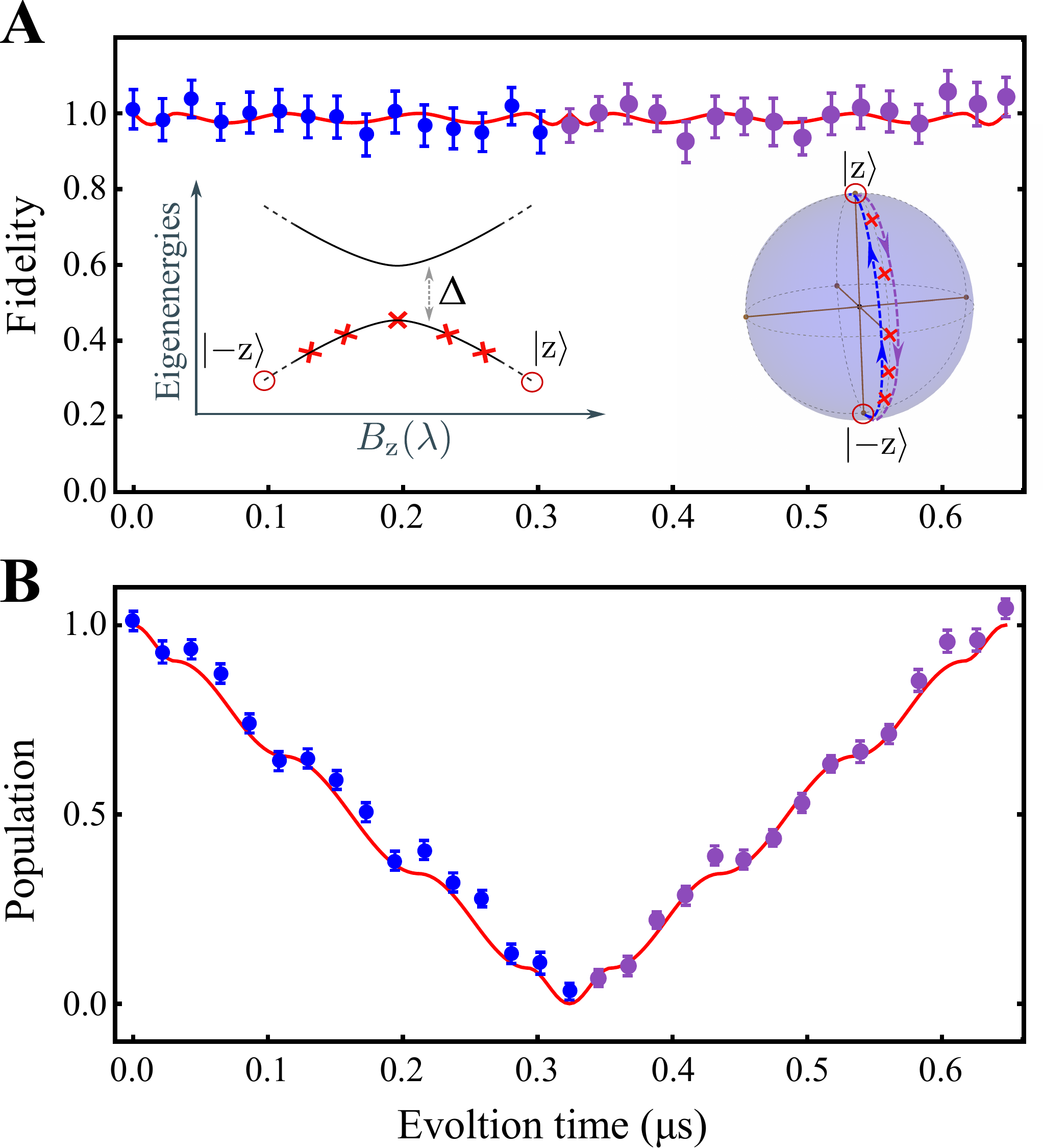}
\caption{\label{fig6:LZ}\textbf{Avoiding unphysical
points in the Landau-Zener model.} (\textbf{A}) Measured fidelity (dots) to the adiabatic state during the evolution time along the path
of the Landau-Zener model by using the jumping protocol of $N=5$ pulses.
Left inset is the eigenenergies of the Landau-Zener model with an avoid crossing $\Delta=2\pi\times 5$ MHz. The path (right inset) was set as from the initial eigenstate $|\rm{-z}\rangle$ to  $|\rm{z}\rangle$ (blue arrows) and back from $|\rm{z}\rangle$ to $|\rm{-z}\rangle$ (purple arrows). The red circles indicate the unphysical path points ($|\pm\rm{z}\rangle$) that require an infinitely large $B_{\rm{z}}$ and eigenenergies in the Hamiltonian. The jumping protocol avoids the use of the unphysical points and adiabatically transfers the states $|\pm\rm{z}\rangle$ with high fidelity (dots) during the evolution, by jumping only on the path points indicated by red crosses.
(\textbf{B}) Population at $|-\rm{z}\rangle$ during the evolution time along the path. The red lines are the numerical simulations, and the target state is defined as the ideal adiabatic state driven under the control with an infinite number of $\pi$ pulses.}
\end{figure}

Without going through all the path points, the jumping protocol has
advantages to avoid path points (i.e., Hamiltonian with certain eigenstates)  that can not be realized in experiments.
As a proof-of-principle experiment, we consider the Landau-Zener (LZ)
Hamiltonian~\cite{Shevchenko2010}
\begin{equation}
H(\lambda)=H_{\rm{LZ}}(\lambda)\equiv B_{\rm{z}}(\lambda)\frac{\sigma_{\rm{z}}}{2}+\Delta\frac{\sigma_{\rm{x}}}{2}, \label{eq:H_LZ}
\end{equation}
with $\sigma_{\alpha}$ ($\alpha=$ x, y, z) being the Pauli matrices.
Because $\Delta$ is non-zero in the LZ Hamiltonian, tuning the system eigenstates to the eigenstates $|\pm{\rm{z}}\rangle$ of $\sigma_{\rm{z}}$
requires $B_{\rm{z}} \rightarrow \pm \infty$.
Therefore for a perfect state transfer from $|\Psi_{\rm{i}}\rangle=|-{\rm{z}}\rangle$
to $|\Psi_{\rm{t}}\rangle=|{\rm{z}}\rangle$ by using the standard continuous protocol,
it is required to adiabatically tune $B_{\rm{z}}$ from $-\infty$
to $+\infty$ (see insets of Fig.~\ref{fig6:LZ}A).
The experimental implementation of $B_{\rm{z}}=\pm\infty$ however requires an infinitely large control field, which is a severe limitation.
In our experiment, a large $B_{\rm{z}}$ field can be simulated by going to the rotating frame of the microwave control field with a large frequency detuning.
The experimental realization of $B_{\rm{z}} \rightarrow \pm \infty$ can be challenging in other quantum platforms.
For example, for superconducting qubits where $\Delta/(2\pi)$ could be as large as 0.1 GHz but
the tuning range of $B_{\rm{z}}/(2\pi)$ is usually limited to a couple of GHz or even of the same order of magnitude as $\Delta/(2\pi)$~\cite{sun2015observation}.  For two-level quantum system comprising Bose-Einstein condensates in optical lattices, the maximum ratio of $B_{\rm{z}}/\Delta$ is determined by the band structure~\cite{bason2012high}. For singlet-triplet qubits in semiconductor quantum dots, the exchange interaction for the control of $B_{\rm{z}}$ is positively confined~\cite{foletti2009universal}.
On the contrary, with the jumping
approach, one can avoid the unphysical points such as $B_{\rm{z}}=\pm\infty$
as infinitely slow and continuous process is not required and achieve high-fidelity state transfer as shown in Fig.~\ref{fig6:LZ}.

As a remark, we find that our jumping protocol with $N=1$ (i.e., a Rabi pulse)
specializes to the optimized composite pulse protocol~\cite{bason2012high}
but has the advantage that no additional strong $\pi/2$ pulses at
the beginning and the end of the evolution are required.
Moreover, by applying the jumping protocol with $N=1$ to the adiabatic passage proposed in \cite{Cirac1994},
we obtain the protocol that has been used to experimentally generate Fock states of a trapped atom~\cite{Meekhof1996}.
When, instead of a single target point, high-fidelity adiabatic evolution along the path is also desired,
we can use the jumping protocol with a larger $N$.

\subsection*{Conclusion and outlook}
In summary, our experiments demonstrated that energy level crossings and vanishing gaps allow and can even accelerate quantum adiabatic evolutions, challenging the traditional view that adiabatic control must be slow and unit-fidelity adiabatic processes require an infinite amount of evolution time. By experimentally verifying a recently derived quantum adiabatic condition, we have shown that the quantum dynamic phases are more fundamental than energy gaps in quantum adiabatic processes. Thanks to rapid changes of these phases, non-adiabatic transitions can be efficiently suppressed and fast varying Hamiltonians can still realize quantum adiabatic evolutions. Our results break the limit imposed by the conventional adiabatic methods which originate from the traditional concept of extremely slow change in classical mechanics~\cite{Laidler1994,Born1928}, allowing fast quantum adiabatic protocols with unit fidelity within finite evolution times. In addition, the freedom of using vanishing gaps provides the ability to avoid unphysical points in an adiabatic path and allows to incorporate pulse techniques~\cite{yang2011preserving} into a quantum adiabatic evolution to suppress environmental noise for long-time robust adiabatic control.
While it is possible to mimic the infinitely slow quantum adiabatic evolution by using additional counterdiabatic control, i.e., shortcuts to adiabaticity~\cite{Demirplak2003,Berry2009,torrontegui2013shortcuts,Deffner2014,Zhou2017,bason2012high}, 
the implementation of the counterdiabatic control can be exceedingly intricate because it may
need interactions absent in the system Hamiltonian~\cite{Deffner2014,Zhou2017}. Furthermore, the counterdiabatic control unavoidably changes the eigenstates of the initial Hamiltonian and introduces additional control errors~\cite{Deffner2014,Zhou2017}. However, because our protocol uses the intrinsic adiabatic path that follows the eigenstates of the Hamiltonian, no additional control is required. As a consequence, our methods avoid the use of difficult or unavailable control resources and share the intrinsic robustness of adiabatic methods. With the removal of the prerequisites in the conventional adiabatic conditions, namely non-zero gaps and slow control, our results provide new directions and promising strategies for fast, robust control on quantum systems.

\section*{Materials and Methods}
\subsection*{Adiabatic evolution along the geodesics of a general quantum system}

For two arbitrary states \wzy{(e.g., entangled states and product states)} of a general quantum system, $|\Psi_{\rm{i}}\rangle$ and $|\Psi_{\rm{t}}\rangle$,
one can write $\langle\Psi_{\rm{i}}|\Psi_{\rm{t}}\rangle=\cos\left(\frac{1}{2}\theta_{\rm{g}}\right)e^{i\phi_{\rm{i,t}}}$
with $\phi_{\rm{i,t}}$ and $\theta_{\rm{g}}$ being real. Here
$\theta_{\rm{g}}$ is the path length connecting $|\Psi_{\rm{i}}\rangle$
and $|\Psi_{\rm{t}}\rangle$ by the geodesic and we set $\phi_{\rm{i,t}}=0$
by a proper gauge transformation~\cite{anandan1990geometry}. The geodesic
\cite{anandan1990geometry,chruscinski2004geometric} that connects $|\Psi_{\rm{i}}\rangle$ and $|\Psi_{\rm{t}}\rangle$ by varying $\lambda=0$ to $\lambda=1$ can be written
as $|\psi_{1}(\lambda)\rangle=c_{\rm{i}}(\lambda)|\Psi_{\rm{i}}\rangle+c_{\rm{t}}(\lambda)|\Psi_{\rm{t}}\rangle$,
where the coefficients $c_{\rm{i}}(\lambda)=\cos(\frac{1}{2}\theta_{\rm{g}}\lambda)-\sin(\frac{1}{2}\theta_{\rm{g}}\lambda)\cot(\frac{1}{2}\theta_{\rm{g}})$
and $c_{\rm{t}}(\lambda)=\sin(\frac{1}{2}\theta_{\rm{g}}\lambda)/\sin(\frac{1}{2}\theta_{\rm{g}})$
for $\sin(\frac{1}{2}\theta_{\rm{g}})\neq0$. To describe $|\psi_{1}(\lambda)\rangle$
in terms of the system eigenstates, we choose an orthonormal state
$|\psi_{2}(0)\rangle\propto\left(I-|\Psi_{\rm{i}}\rangle\langle\Psi_{\rm{i}}|\right)|\Psi_{\rm{t}}\rangle$
if $\sin(\frac{1}{2}\theta_{\rm{g}})\neq0$. When $|\Psi_{\rm{t}}\rangle$
is equivalent to $|\Psi_{\rm{i}}\rangle$ up to a phase factor (i.e., $\sin(\frac{1}{2}\theta_{\rm{g}})=0$), $|\psi_{2}(0)\rangle$ can be an arbitrary orthonormal state. Then the geodesic can be written
as $|\psi_{1}(\lambda)\rangle=\cos(\frac{1}{2}\theta_{\rm{g}}\lambda)|\psi_{1}(0)\rangle+\sin(\frac{1}{2}\theta_{\rm{g}}\lambda)|\psi_{2}(0)\rangle$
and its orthonormal state $|\psi_{2}(\lambda)\rangle=-\sin(\frac{1}{2}\theta_{\rm{g}}\lambda)|\psi_{1}(0)\rangle+\cos(\frac{1}{2}\theta_{\rm{g}}\lambda)|\psi_{2}(0)\rangle$.
Along the geodesic we have $g_{2,1}(\lambda)=-g_{1,2}(\lambda)=i\frac{1}{2}\theta_{\rm{g}}$
being a constant and $g_{n,m}=0$ for other combinations of $n$ and $m$.
Along the geodesic if one changes the dynamic phases with
a $\pi$ phase shift only at each of the $N$ equally spaced
path points $\lambda_{j}=(2N)^{-1}(2j-1)$ with $j=1,2,\ldots,N$, the operators $W(\lambda)$ at different
$\lambda$ commute and we have $\int_{0}^{1}e^{i\phi_{1,2}}d\lambda=0$. As a consequence,
$U_{\rm{dia}}(1)=\exp\left[i\int_{0}^{1}W(\lambda)d\lambda\right]=I$
and the quantum evolution $U=U_{\rm{adia}}$ does not have any non-adiabatic
effects.

\subsection*{Hamiltonian of the NV center under microwave control}
Under a magnetic field $b_{\rm{z}}$ along the NV
symmetry axis, the Hamiltonian of the NV center electron spin without microwave control reads
$H_{\rm{NV}}=D S_{\rm{z}}^{2}-\gamma_{\rm{e}}b_{\rm{z}}S_{\rm{z}}$,
where $S_{\rm{z}}$ is the electron spin operator, $D\approx2\pi\times2.87$ GHz is the ground
state zero field splitting, $b_{\rm{z}}$ is the magnetic field, and $\gamma_{\rm{e}}=-2\pi\times 2.8$ MHz G$^{-1}$ is the electron spin gyromagnetic
ratio~\cite{doherty2013nitrogen}.

Following the standard methods to achieve a controllable Hamiltonian for quantum adiabatic evolution~\cite{RMP2017,Vitanov2001},
we apply a microwave field $\sqrt{2}\Omega(\lambda)[\cos(\omega_{\rm{mw}} t + \vartheta(\lambda)]$ to the NV $m_{\rm{s}}=0$ and $m_{\rm{s}}=1$ levels to form a qubit with the qubit states $|\rm{z}\rangle\equiv |m_{\rm{s}}= 1\rangle$ and $|-\rm{z}\rangle\equiv |m_{\rm{s}}=0\rangle$. The microwave frequency $\omega_{\rm{mw}}$ may also be tuned by the parameter $\lambda$ to realize a controllable frequency detuning $\delta(\lambda)$ with respect to the transition frequency of $m_{\rm{s}}=0$ and $m_{\rm{s}}=1$ levels. In the standard rotating frame of the microwave control field, we have
the general qubit Hamiltonian under the microwave control~\cite{doherty2013nitrogen}
\begin{equation}
H(\lambda) = \delta(\lambda) \frac{\sigma_{\rm{z}}}{2} + \Omega(\lambda)\left[\cos\vartheta(\lambda)\frac{\sigma_{\rm{x}}}{2} +\sin\vartheta(\lambda)\frac{\sigma_{\rm{y}}}{2}\right], \label{eq:HLambda}
\end{equation}
where the microwave phase $\vartheta(\lambda)$, microwave detuning $\delta(\lambda)$, and microwave Rabi frequency $\Omega(\lambda)$ are all tunable and can be time-dependent in experiment.
The usual Pauli operators satisfy $\sigma_{\rm{z}}{|\pm \rm{z}\rangle}=\pm {|\pm \rm{z}\rangle}$ and $\left[\cos(\theta_{\rm{g}}\lambda)\sigma_{\rm{x}} +\sin(\theta_{\rm{g}}\lambda)\sigma_{\rm{y}}\right]|\pm_{\lambda}\rangle=\pm |\pm_{\lambda}\rangle$, where the states $|\pm_{\lambda}\rangle$ are given by Eq.~\ref{eq:xyPath}.

By setting the microwave detuning $\delta(\lambda)=0$, we achieve the Hamiltonian in Eq.~\ref{eq:Hxy}, which in terms of the Pauli operators reads
$$H_{\rm{XY}}(\lambda) = \Omega(\lambda)\left[\cos(\theta_{\rm{g}}\lambda)\frac{\sigma_{\rm{x}}}{2} +\sin(\theta_{\rm{g}}\lambda)\frac{\sigma_{\rm{y}}}{2}\right].$$
By varying the parameter $\lambda$, the system eigenstates follow the geodesics along the equator of the Bloch sphere where the north and south poles are defined by the states $|\pm\rm{z}\rangle$. Here the energy gap $\Omega(\lambda)$ is directly controlled by the amplitude of the microwave field.

On the other hand, by using a constant Rabi frequency $\Omega(\lambda)=\Delta$ and a tunable frequency detuning $\delta(\lambda)=B_{\rm{z}}(\lambda)$, we obtain the Landau-Zener Hamiltonian $H_{\rm{LZ}}(\lambda)$ given by Eq.~\ref{eq:H_LZ}.

\subsection*{Adiabatic evolution by continuous driving with a constant gap}
Consider a conventional adiabatic driving that a constant amplitude driving field rotates around the z axis, with the Hamiltonian 
$H(\lambda)=\frac{1}{2}\Omega e^{-i\frac{1}{2}\sigma_{z}\theta_{\rm{g}}\lambda}\sigma_{\theta}e^{i\frac{1}{2}\sigma_{z}\theta_{\rm{g}}\lambda}$, which
is parameterized by $\lambda=t/T$ along a circle of latitude with $\sigma_{\theta}=\sigma_{z}\cos\theta+\sigma_{x}\sin\theta$
in a total time $T$. The difference of the accumulated dynamic phases
at $\lambda=1$ on the two eigenstates is $\phi=\Omega T$. One can show
that the system evolution at $\lambda=1$ reads
\begin{equation}
U=e^{-i\frac{1}{2}\theta_{\rm{g}}\sigma_{z}}\exp\left[-i\frac{1}{2}\left(\phi\sigma_{\theta}-\theta_{\rm{g}}\sigma_{z}\right)\right].\label{eq:Uconstant}
\end{equation}
The ideal adiabatic evolution is obtained by using Eq.~\ref{eq:Uconstant}
in the adiabatic limit $T\rightarrow\infty$ (i.e., $\phi\rightarrow\infty$),
$$U_{\rm{adia}} =\lim_{T\rightarrow\infty}U=e^{-i\frac{1}{2}\theta_{\rm{g}}\sigma_{z}}e^{i\frac{1}{2}\theta_{\rm{g}}\cos\theta\sigma_{\theta}}e^{-i\frac{1}{2}\phi\sigma_{\theta}}.$$
Without the part of dynamic phases, $U_{\rm{adia}}$ describes
geometric evolution and for a cyclic evolution (i.e., $\theta_{\rm{g}}=2\pi$)
the geometric evolution is described by the Berry's phases $\pm\pi(\cos\theta-1)$.
By comparing $U_{\rm{adia}}$ and $U$ or by using the results of \cite{wang2016necessary},
the non-adiabatic correction is given by
\begin{equation}
U_{\rm{dia}}=\exp\left[i\frac{1}{2}\left(\phi-\theta_{\rm{g}}\cos\theta\right)\sigma_{\theta}\right]U^{\prime},\label{eq:UDia_Const}
\end{equation}
with
$$U^{\prime}=\exp\left[-i\frac{1}{2}\left(\phi\sigma_{\theta}-\theta_{\rm{g}}\sigma_{z}\right)\right].
$$
In the adiabatic limit $T\rightarrow\infty$ (i.e., $\phi\rightarrow\infty$), $U_{\rm{dia}}=I$
is the identify operator. We note that when the phase factor of the state is irrelevant
one can perform perfect state transfer by this driving if the initial state is prepared in an initial
eigenstate of the driving Hamiltonian $H(\lambda)$ (i.e., an eigenstates
of $\sigma_{\theta}$).
From Eq.~\ref{eq:UDia_Const}, $U_{\rm{dia}}$ is diagonal
in the basis of $\sigma_{\theta}$ when $U^{\prime}\propto I$.
As a consequence, when $U^{\prime}\propto I$ and $|\Psi_{\rm{i}}\rangle$ is prepared as an eigenstate
of $\sigma_{\theta}$ [and hence $H(\lambda=0)$], the evolved state $U|\Psi_{\rm{i}}\rangle$
matches the target state $U_{\rm{adia}}|\Psi_{\rm{i}}\rangle$
up to a phase factor. For the case of the evolution along the geodesic (e.g., $\theta=\pi/2$) and $\sqrt{\phi^{2}+\theta_{\rm{g}}^{2}}=2k\pi$
($k=1,2,\ldots$), we have $U^{\prime}\propto I$ and therefore the population transfer for the initial eigenstates
of $\sigma_{x}$ is perfect.

\subsection*{Numerical simulations}

In the simulations, we modelled dephasing noise and random fluctuations by adding them to
the Hamiltonian Eq.~\ref{eq:HLambda} via $\delta(\lambda)\rightarrow \delta(\lambda)+ \delta_0$ and $\Omega(\lambda)\rightarrow \Omega(\lambda)\wzy{(1 + \delta_1)}$.
Here $\delta_0$ is the dephasing noise from static and time-dependent magnetic
field fluctuations with a $T_{2}^{*}=1.7$ $\mu$s. $\delta_1$ is the random static
changes in the driving amplitude.  $\delta_0$ follows the Gaussian distribution with the mean value $\mu=0$ and the standard deviation $\sigma=2\pi\times 130$ kHz. The probability density of $\delta_1$ has the Lorentz form $f(\delta_1,\gamma)=1/\{\pi\gamma[1+(\wzy{\delta_1}/\gamma)^2]\}$ with $\gamma=0.0067$. All the parameters in the distribution function are extracted from fitting the free induction decay (FID) and the decay of Rabi oscillation.

\subsection*{Experimental Setup}

The experiments were performed with a home-built optically detected magnetic resonance (ODMR) platform, which consists of a confocal microscope and a microwave (MW) synthesizer
%(fig.~\ref{fig:FigSetup})
(fig.~S5). A solid state green laser with 532 nm wavelength is used for initializing and reading out the NV spin state. The light beam was focus on the NV center through an oil immersion objective (N.A., 1.4). The emitted fluorescence from NV center was collected by a single photon counting module (APD).  Here we used an NV center embedded in a room-temperature bulk diamond grown by chemical vapor deposition with [100] faces. It has $^{13}$C isotope of
natural abundance and nitrogen impurity less than 5 ppb. To lift the degeneracy of the $|m_{\rm{s}}=\pm1\rangle$ states, a static magnetic field of 510 G was provided by a permanent magnet. The magnetic field was aligned  by adjusting the three-dimensional positioning  stage on which the magnet was mounted, and simultaneously monitoring the counts of the NV center. The direction of the magnetic field is well aligned when the counts show no difference between with and without the magnet. Manipulation of the NV center is performed by MW pulses applied through a home-made coplanar waveguide (CPW). The MW pulses were generated by the I/Q modulation of the Agilent arbitrary wave generator (AWG) 81180A and the vector signal generator (VSG) E8267D and then amplified by Mini Circuits ZHL-30W-252+. An atomic clock was used to synchronize the timing of the two. The AWG supplies the I and Q data with a frequency of 400 MHz, and the VSG generates the 3898 MHz carrier. The output frequency is 4298 MHz, which matches the transition frequency between the NV $m_{\rm{s}}=0$ and $m_{\rm{s}}=+1$ states.

\subsection*{Experimental Sequences}

As the magnetic field is 510 G, we first applied the green laser for 3~$\mu$s to initialize the NV center electronic spin to the level of $m_{\rm{s}}=0$ and to polarize the adjacent $^{14}$N nuclear spin simultaneously~\cite{Epstein2005}.
The preparation of the NV electron spin in an equal superposition state of $m_{\rm{s}}=0$ and $m_{\rm{s}}=1$ was realized by applying a MW $\pi_{\rm{x}}/2$ ($\pi_{\rm{y}}/2$) pulse, i.e., by the rotation around the x (y) axis with an angle of ${\pi}/{2}$.
Then the NV electron spin was driven according to a desired path. To experimentally
characterize the evolution path, we sampled the path with several points
and measured the spin state through tomography. ${\pi_{\rm{x}}}/{2}$ or ${\pi_{\rm{y}}}/{2}$  pulses were applied to readout the off-diagonal terms.
Finally the spin state was read out by applying
the laser pulse again and measuring the spin-dependant fluorescence.
Typically the whole sequence was repeated $10^{5}$ times to get a better signal to noise ratio (SNR). The schematic diagram of the pulse sequence is shown in fig.~S6.
%fig.~\ref{fig:smFigSequence}.

In driving the NV electron spin along the path given by Eq.~\ref{eq:xyPath}, we used an on-resonant MW field and swept the MW phase $\theta_{\rm{g}}\lambda$ with the path parameter $\lambda$. In driving the NV electron spin along the path of the LZ Hamiltonian (see Eq.~\ref{eq:H_LZ}), the MW phase was a constant, $\Delta$ was set by the Rabi frequency, and $B_{\rm{z}}$ was the MW frequency detuning which varied as $B_{\rm{z}}=-\Delta\cot(\theta_{\rm{g}}\lambda)$.
For continuous driving, the path parameter $\lambda$ varies with a constant rate $d\lambda/dt=f_{\rm{rot}}$.
In the jumping protocol $\lambda$ jumps from point to point: $\lambda=\lambda_{j}=(2N)^{-1}(2j-1)$ for $j=1,2,\ldots,N$. In this work the jumping protocol had a constant driving Rabi frequency $\Omega_{0}$ and $\lambda= \lambda_j$ if $(j-1)T/N\leq t<j T/N$ for a path with $N$ pulses applied in a total time $T$.
In the experiments with the back-forward motion along the geodesic, we reversed
the order of the parameter $\lambda$ in the backward path. That is, in the jumping protocol  we repeat the subsequent parameters $(\lambda_{1}, \lambda_{2},\ldots, \lambda_{N-1},\lambda_{N}, \lambda_{N}, \lambda_{N-1}, \ldots, \lambda_{2}, \lambda_{1})$, while for the standard
protocol of continuous driving we used the rate $d\lambda/dt=f_{\rm{rot}}$ for a forward path
and the rate  $d\lambda/dt=-f_{\rm{rot}}$ for a backward path and repeated the process.

We removed the irrelevant dynamic phases if the initial state was not prepared in an initial eigenstate to reveal the geometric evolution. At the beginning of state readout, we compensated the dynamic phases by applying an additional driving with a microwave $\pi$ phase shift (i.e., $\Omega\rightarrow-\Omega$) at the point of the target state for a time equalling to the time for adiabatic evolution. This additional driving did not change the geometric phases and state transfer because it was applied at the final path point.

% Your references go at the end of the main text, and before the
% figures.  For this document we've used BibTeX, the .bib file
% scibib.bib, and the .bst file Science.bst.  The package scicite.sty
% was included to format the reference numbers according to *Science*
% style.

%\bibliography{scibib}
\bibliographystyle{Science}

\clearpage

\section*{Acknowledgements}
{\bf Funding:} The authors at USTC are supported by the National Key Research and Development Program of China (Grant No.~2018YFA0306600, No.~2013CB921800, No.~2016YFA0502400 and No.~2017YFA0305000), the NNSFC (Grants No.~81788101, No.~11227901, No.~11722544, No.~91636217 and No.~11775209), the CAS (Grants No. GJJSTD20170001 and No. QYZDY-SSW-SLH004), Anhui Initiative in Quantum Information Technologies (Grant No. AHY050000), the CEBioM , the IPDFHCPST (Grant No.~2017FXCX005)\wzy{, the Fundamental Research Funds for the Central Universities,} and the Thousand Young Talents Program. The authors of Ulm University are supported by the ERC Synergy grant BioQ, \wzy{the EU projects HYPERDIAMOND and AsteriQs, the BMBF projects NanoSpin, DiaPol} and via the IQST which is financially supported by the Ministry of Science, Research and Arts Baden W{\"u}rttemberg.
{\bf Author contributions:} J.D. supervised the entire experiment. Z.Y.W. and M.B.P. formulated the theory. J.D. and F.S. designed the experiments. K.X., X.X., and P.W. prepared the setup. K.X. and T.X. performed the experiment. T.X., Z.Y.W., and K.X. performed the simulation. Z.Y.W., K.X., T.X., M.B.P., Y.W., and J.D. wrote the manuscript. All authors discussed the results and commented on the manuscript.
{\bf Competing interests:} All authors declare no competing financial interests.
{\bf Data and materials availability: } All data is available in the main text or the supplementary materials.

\end{document}